\documentclass[aps,prb,reprint,superscriptaddress]{revtex4-1}

\usepackage{graphicx}
\usepackage{color}

\begin{document}

\title{Electronic structure and magnetic anisotropies of
antiferromagnetic transition-metal difluorides}

\author{Cinthia Antunes Corr\^ea}
\affiliation{Institute of Physics, Academy of Science of the Czech Republic, 
Cukrovarnick\'a 10, Praha 6, Czech Republic}
\affiliation{Department of Physics of Materials, Charles University in Prague,
  Ke Karlovu 5, 121 16 Prague, Czech Republic}
\author{Karel V\'yborn\'y}
\affiliation{Institute of Physics, Academy of Science of the Czech Republic, 
Cukrovarnick\'a 10, Praha 6, Czech Republic}

\begin{abstract}
We compare GGA+U calculations with available experimental data
and analyze the origin of magnetic anisotropies in MnF$_2$, FeF$_2$,
CoF$_2$, and NiF$_2$.  We confirm that the magnetic anisotropy of MnF$_2$
stems almost completely from the dipolar interaction, while
magnetocrystalline anisotropy energy plays a dominant role in the other
three compounds, and discuss how it depends on the details of band
structure. The last mentioned is critically compared to available
optical measurements. The case of CoF$_2$, where magnetocrystalline
anisotropy energy (MCA) strongly depends on $U$, is put into
contrast with FeF$_2$ where theoretical predictions of magnetic
anisotropies are nearly quantitative. 
\end{abstract}

\pacs{later}

\maketitle

\section{Introduction}

Several rutile-structure difluorides of transition metals (TMs) such
as MnF$_2$ have long been known to be antiferromagnetically ordered at
low temperatures. Albeit not the first antiferromagnets (AFMs) ever
identified, they received significant attention in the late '50s
and '60s when their magnetic anisotropy was effectively determined
using measurements of spin flop. These materials are arguably one of
the simplest AFMs one can imagine: their two magnetic sublattices are
oriented in opposite directions (collinearity) and they exhibit
uniaxial MA, which reduces the complexity of domain
building.  Renewed interest in these materials
has arisen recently in the context of antiferromagnetic
spintronics \cite{Baltz:2017_a}. Very recent device concepts using
these traditional AFMs include bilayers where spin pumping by
AFM \cite{Johansen:2017_a} or spin Seebeck effect \cite{Rezende:2016_a}
could be observed \cite{Wu:2016_a}. Spin currents can be passed through
insulating AFMs \cite{Moriyama:2015_a} and devices involving MnF$_2$
and FeF$_2$ have been suggested \cite{Johansen:2017_a,Gulbrandsen:2018_a}. 

Motivated initially by the lack of theoretical estimates of
magnetocrystalline anisotropy (MCA), we soon realized that not even the band
structures of magnanese, iron, cobalt and nickel difluorides are well
established in the literature. We therefore present DFT+U calculations
(described in detail in Appendix~C), 
compare them to optical measurements where available, identify the
missing information (and propose experiments and calculations to be
still carried out) and finally present the MCA calculations and
discuss their agreement with experimentally determined magnetic
anisotropies of these materials.

For certain purposes, {\em simple} (in the sense explained above),
AFMs can be described by Stoner-Wohlfarth
model \cite{Bogdanov:2007_a} where the
energy (per volume) divided by sublattice magnetization $M$ reads
\begin{equation}\label{eq-01}
  \frac{E}{MV}= B_e\vec{m}_1\cdot\vec{m}_2 -B\vec{b}\cdot(\vec{m}_1+\vec{m}_2)
  -B_a [(\vec{m}_1\cdot \hat{z})^2+(\vec{m}_2\cdot \hat{z})^2].
\end{equation}
Here, $\vec{m}_{1,2}$ and $\vec{b}$ are the unit vectors giving the
direction of sublattice magnetizations and magnetic field $B$,
respectively. The two material-specific parameters of this model are
the exchange field $B_e$ and anisotropy field $B_a$ and, typically,
$B_e\gg B_a, B$. For $\vec{b}||\hat{z}$, model Eq.~(\ref{eq-01}) implies a
spin flop at $B=B_{sf}=2\sqrt{B_aB_e}$, i.e., abrupt ground state
transition from $\vec{m}_{1,2}||\hat{z}$ with $\vec{m}_{1,2}$ strictly
antiparallel to, approximately $\vec{m}_{1,2}\perp\hat{z}$ with
$\vec{m}_{1,2}$ slightly canted (see Fig.~\ref{fig-07} in
Appendix~B). Using this effect, $B_a$ can be determined from
magnetometry provided that the exchange field is known or estimated.

\begin{table*}
\centering
\small
\caption{Parameters of MnF$_2$, FeF$_2$, CoF$_2$, and NiF$_2$ related
  to magnetism. Note that definitions of $B_e$ and $B_a$ vary
  throughout literature.}
\begin{tabular}{lp{2.5cm}l|p{2.5cm}l|p{2.cm}l|p{2.cm}l} \\ \hline

            &\multicolumn{2}{c}{MnF$_2$}     &\multicolumn{2}{c}{FeF$_2$}       
               &\multicolumn{2}{c}{CoF$_2$}         &\multicolumn{2}{c}{NiF$_2$}
        \\ \hline

            & exp                       &calc& exp                         & calc              & exp                          &calc & exp        & calc              \\

mag.mom. [$\mu_B$]&5.04 \cite{Strempfer:2004_a}&4.4&3.93 \cite{Strempfer:2004_a}, 3.75 \cite{Novak:2006_a}&3.6&2.21 \cite{Strempfer:2004_a}&2.6 &1.96 \cite{Strempfer:2004_a}&1.63 \\

ideal S           &2.5  &    &2      &            &1.5     &     &1 &            \\ \hline

$B_e$ [T]   &46.5 \cite{Carrico:1994_a}, 57.5 \cite{Rezende:2016_a}&85.5&43.4 \cite{Carrico:1994_a}, 62 \cite{Rezende:2016_a}&116.7&32.4 \cite{Carrico:1994_a}&67.4&   &163.5                \\

$B_a$ [T]   &0.697 \cite{Carrico:1994_a}, 0.8 \cite{Rezende:2016_a}&0.42&14.9 \cite{Carrico:1994_a}, 19.2 \cite{Rezende:2016_a}&2.6&3.2 \cite{Carrico:1994_a}&0.73$^*$&& $-0.50$      \\
$B_a^{(1)}$ [T]  &&  0.2$\cdot10^{-3}$ && 2.3 && 0.52$^*$ && $-0.71$\\
dipolar term&                          & 418 mT&                             & 317 mT           &                              & 211 mT &           &      203 mT             \\ \hline

$B_{sf}$ [T]&9.27 \cite{Jacobs:1961_a}                  & 12.0 &41.9 \cite{Jaccarino:1983_a}     & 34.8              &14.0 \cite{Carrico:1994_a}                          & ***  &           &                   \\

$T_N$ [K]   &67.7 \cite{Strempfer:2004_a}&&75.8 \cite{Strempfer:2004_a}&&37.7 \cite{Nagamiya:1955_a}&&74.1 \cite{Strempfer:2004_a}& \\             
\end{tabular}
\label{tab-01}
\end{table*}

We summarize the measured values of $B_{sf}$ for the first three
compounds of the series in Tab.~\ref{tab-01} and compare them to
theoretically calculated values. The latter are obtained by combining
$B_a$, which comprises MCA and dipolar interaction (the former,
$B_a^{(1)}$, calculated by \textit{ab initio} methods detailed in
Sec.~III), and $B_e$, based on an estimate of the exchange coupling $J$
from the N\'eel temperature \cite{explEq2},
\begin{equation}\label{eq-02}
  \frac{kT_N}J = \frac13 S(S+1)
\end{equation}
where $S\mu_B$ is the TM atom magnetic moment (relation between
the exchange coupling $J$ and $B_e$ is given in Sec.~III).
We find a satisfactory agreement between experimental and theoretical
values of spin-flop fields for MnF$_2$ and FeF$_2$ and the following
conclusions can be made. Magnetic anisotropy of MnF$_2$ is primarily
driven by dipole interactions (see Appendix A), which is not surprising
given the atomic configuration of manganese ($L=0$ orbital singlet),
which does not allow any appreciable MCA (see comments\cite{sim-MnF2}
on single-ion model in Sec.~III). On the other hand, this does not apply to
FeF$_2$, where the TM $3d$ shell is not half-filled and sizable
matrix elements of $\vec{L}\cdot\vec{S}$ then lead to a strong
magnetocrystalline anisotropy, which translates into spin-flop fields
as large as 42~T.

Calculations of MCA in CoF$_2$ yield ambiguous results (see Sec.~III)
and we, therefore, use opposite reasoning for this material:
using $B_{sf}$ and $B_e$, we estimate $B_a$, which is then shown
to imply $B_a^{(1)}$ consistent with our \textit{ab initio} calculation.
Again, this consistency check is explained in Sec.~III and in
Tab.~\ref{tab-01}, the values of $B_a, B_a^{(1)}$ are marked with an
asterisk to indicate that they are calculated using experimental $B_{sf}$.
Regarding NiF$_2$, we find negative $B_a$ in agreement with
experimental evidence \cite{Shi:2004_a} of $\vec{m}_{1,2}$ oriented in plane.
Spin-flop measurements are more complicated in this case since there
are multiple easy axes (such a system is prone to build multidomain
states) and therefore no data is given for $B_{sf}$ in Tab.~\ref{tab-01}.

Theoretical calculation of MCA relies on a solid knowledge of the band-structure.
 It is essentially the difference between two large numbers
$E_\parallel$ and $E_\perp$, the total energy of the occupied electron
states for $\vec{m}_{1,2}||\hat{z}$ and $\vec{m}_{1,2}\perp\hat{z}$, so
that even small inaccuracies may lead to completely wrong results
unless such inaccuracies accurately cancel (i.e., any error in the
 band-structure determination has the same effect on both $E_\parallel$ and
$E_\perp$). It should be pointed out that these calculations must
include the effect of spin-orbit interaction without which the MCA vanishes
($E_\parallel=E_\perp$). Band structures of the four compounds
considered in this article have been calculated previously under
various approximations:  LSDA band structures of MnF$_2$ and NiF$_2$
were first calculated by Dufek, Schwarz, and Blaha\cite{Dufek:1993_a}
and, a little later, the same group also added FeF$_2$ and CoF$_2$ using
GGA \cite{Dufek:1994_a} (see also Appendix~C), albeit with
unrealistically small gaps. This improved with the advent of
GGA+U \cite{Lopez-Moreno:2012_a,Barreda-Argueso:2013_a} where, however,
not much attention was paid to how large the values of the model parameters
$U,J$ actually should be. Unrestricted Hartree-Fock
calculations \cite{Valerio:1995_a} produce optical gap in excess of 10~eV for
FeF$_2$, which is beyond any doubt too large, realistic size being
close to 3~eV (see below). We now proceed to a discussion of band
structures calculated using GGA+U (based on the same package as in
Ref.~[\onlinecite{Dufek:1994_a}]) and critical comparison of these to
experimentally accessible quantities such as band gap, TM magnetic
moment, and lattice parameters.

\section{Electronic structure}

Given the identical crystal structure (Fig.~\ref{fig-08}) and the
position of Mn, Fe, Co, and Ni in the periodic table, it is not surprising
that structures of all four difluorides are mutually similar. It can
explained using the sketch in Fig.~\ref{fig-01} (see also Fig.~\ref{fig-07}
in Appendix~C). Fermi level ($E_F$) lies in the middle of TM $d$-state
bands and other atomic orbitals (such as fluorine $p$-states) are
relatively far away. The gap that opens within the TM $d$-state band
is partly due to electron-electron interactions (EEIs), which we
model, within density functional theory, by GGA+U (see Appendix~C)
and partly (in the case of Mn and Ni)
due to crystal field effects. Surprisingly, the size of band gaps at
low temperature (i.e., in the AFM phase) is nowhere to be found in the
literature and we can therefore use only some indirect arguments to
support the actual band structure calculations in Fig.~\ref{fig-02}.

\begin{figure}
\includegraphics[scale=0.3]{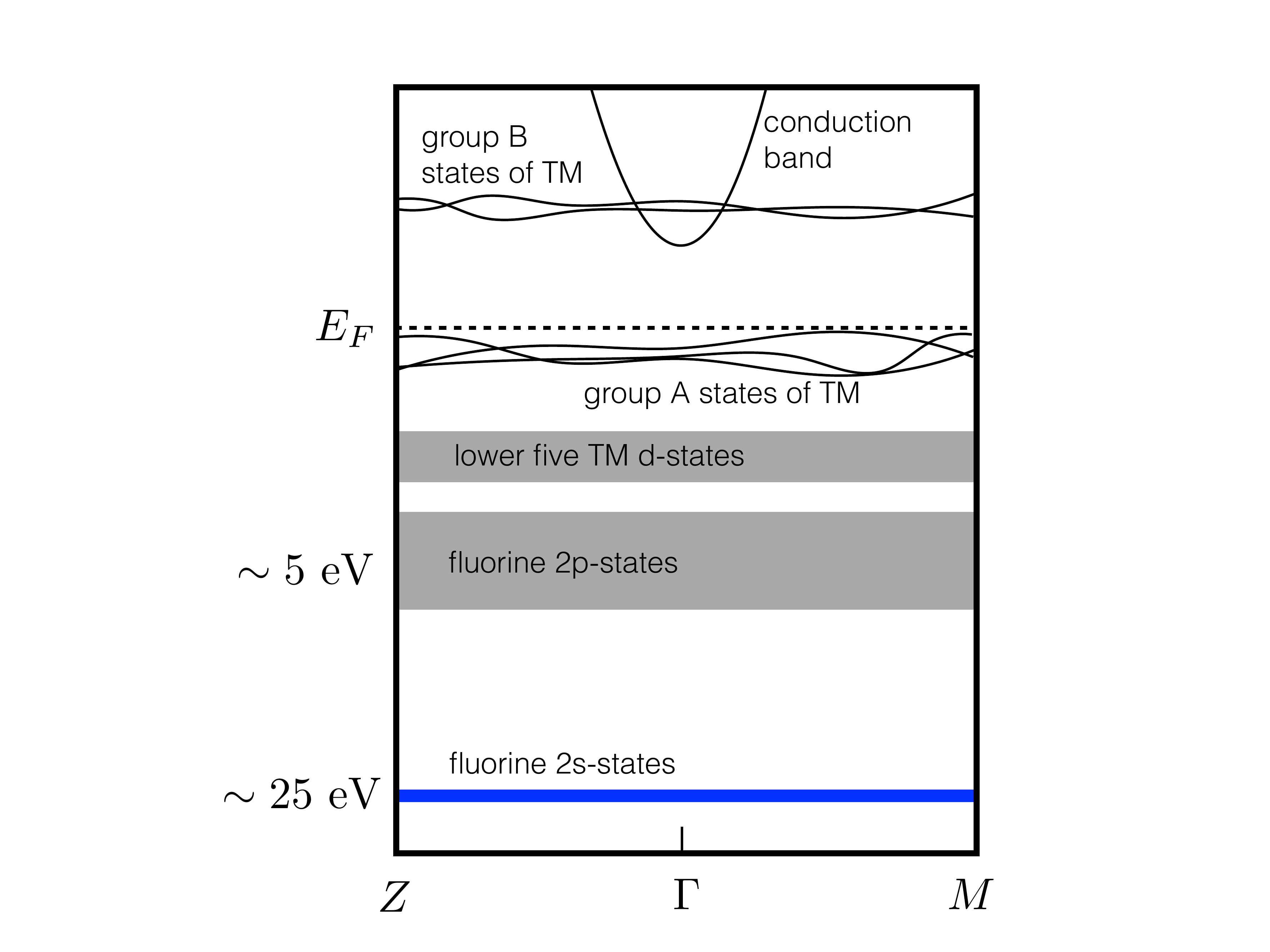}
\caption{Schematic band structure of rutile-type MnF$_2$, FeF$_2$, CoF$_2$, and
  NiF$_2$ in their AFM state. Spin up and down bands are
  degenerate. Note that for MnF$_2$, all five upper $d$-bands are in
  group B (group A is an empty set).}
\label{fig-01}
\end{figure}

\begin{figure*}
\begin{center}
\includegraphics[scale=0.5]{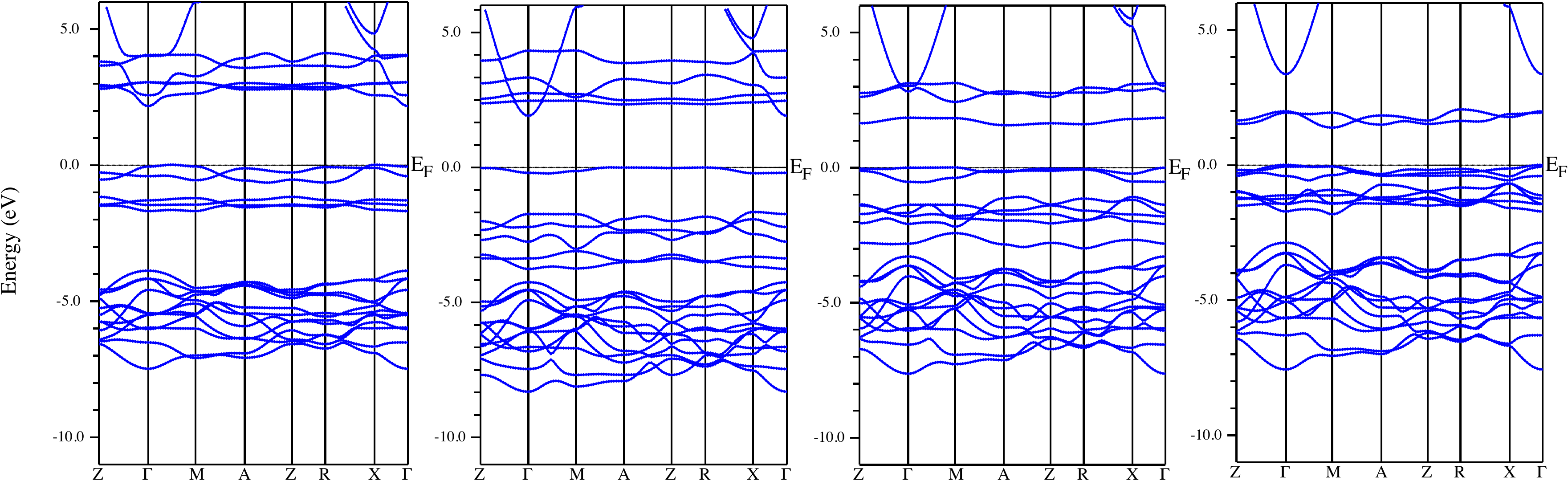}
\end{center}
\caption{Overview of all four fluorides. Left to right: 
MnF$_2$ without $U$,  FeF$_2$ with $U=0.2$~Ry, CoF$_2$ with $U=0.1$~Ry,
and NiF$_2$ again without $U$.}
\label{fig-02}
\end{figure*}

The band structures for spin up and down are the same --- this is a
consequence of the {\em simple} antiferromagnetic order 
(see Fig.~1 in Ref.~[\onlinecite{Baltz:2017_a}] for explanation). 
Focusing on one spin, the total of ten $d$-orbitals (for two TM atoms
in the unit cell, see also Fig.~\ref{fig-08}) divides first into two
quintuplets that could be thought of as of bonding and antibonding
orbitals. The lower quintuplet always lies
below $E_F$, the higher is either partly or completely above
$E_F$. Starting with MnF$_2$ in its $3d^5$ configuration, $E_F$ is
located at the top of the lower five TM $d$-bands and all other five
bands are high above ($\approx 3$~eV or more); in terms of the sketch
in Fig.~\ref{fig-01}, there are no bands in group A and all five are
in group B as the leftmost panel in Fig.~\ref{fig-02}
shows. Another highly dispersive band, which we call ``conduction band''
in Fig.~\ref{fig-01}, crosses these relatively flat $d$-bands (we
comment on this band in Appendix~C).  Now the effect of adding Hubbard $U$
is to push the bands in group B away from the ``lower five'' so that,
since $E_F$ remains pinned at the top of the latter, the group B bands
move higher into the conduction band (compare also the leftmost panel
of Fig.~\ref{fig-02} to Fig.~\ref{fig-07} in Appendix~C).
In other words, the band structure of MnF$_2$ does not change
substantially when $U$ is increased even though the gap does increase
slightly; the gap occurs mainly due to the splitting between the two
quintuplets of $d$-bands and would be present even in the absence of
correlations. However, the effect of Hubbard $U$ is much more dramatic
for the other compounds under scrutiny. We do not discuss the best
choice \cite{explMMMnF2} of $U$ in MnF$_2$ any further since, by virtue
of the argument of half-filled $d$-shell, the MCA is anyway small in
this material.

\begin{figure}
\includegraphics[scale=0.3]{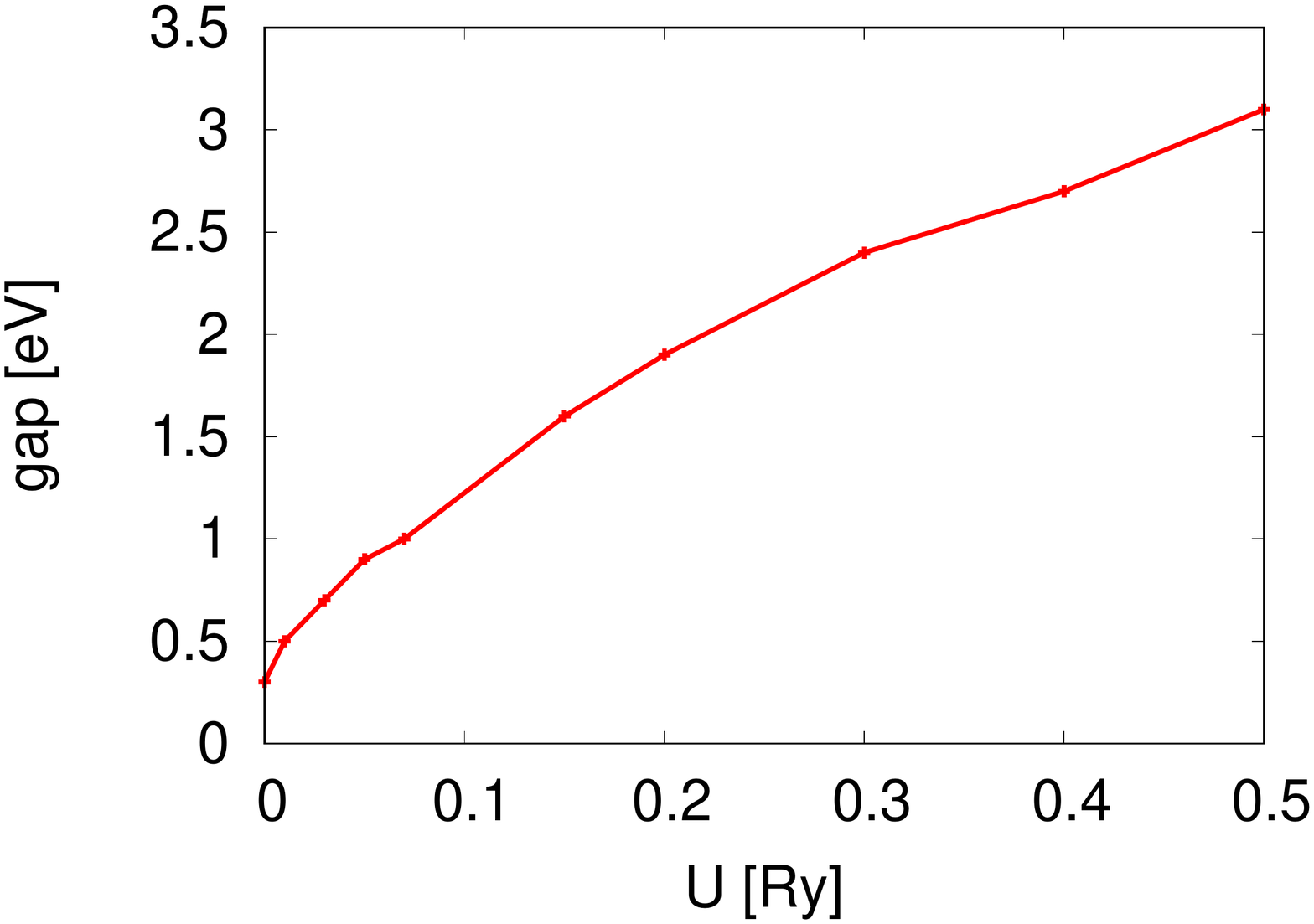}
\caption{Nominal gap in FeF$_2$ as a function of $U$. Note that the
  apparent optical gap is larger because some optical transitions may
  be suppressed.}
\label{fig-03}
\end{figure}

There is one more occupied band in FeF$_2$ than in MnF$_2$ and
therefore one band from group B (Fig.~\ref{fig-01}) has to be
transferred into group A. Because all five $d$-bands are very close
to one another, forming some kind of local spaghetti in the band
structure, this would render FeF$_2$ metallic (at least on the LDA level).
 A better treatment of EEIs is needed. In fact, a gap opens
already by switching to GGA but its size is unrealistically small
($\lesssim 0.5$~eV). Figure~\ref{fig-03} shows that within GGA+U, the
gap grows with $U$ and for values used typically in
literature \cite{Lopez-Moreno:2012_a}, it reaches a
reasonable \cite{Novak:2006_a} size of $\approx 3$~eV. We point out
that a {\em room temperature} measurement of optical
absorption \cite{Santos-Ortiz:2013_a} leads to a similar gap size;
however, we will discuss below the plausibility of using $U$ around
2.5~eV which is somewhat smaller than usual \cite{Thunstrom:2012_a}.
Our choice of $U$ corresponds to the second panel from the left in
Fig.~\ref{fig-02} and seems to give optical spectra closer to another
experimental work on FeF$_2$. Impact of the choice of $U$ on MCA will be
discussed in Sec.~III.

\begin{figure*}
  \includegraphics[scale=0.08]{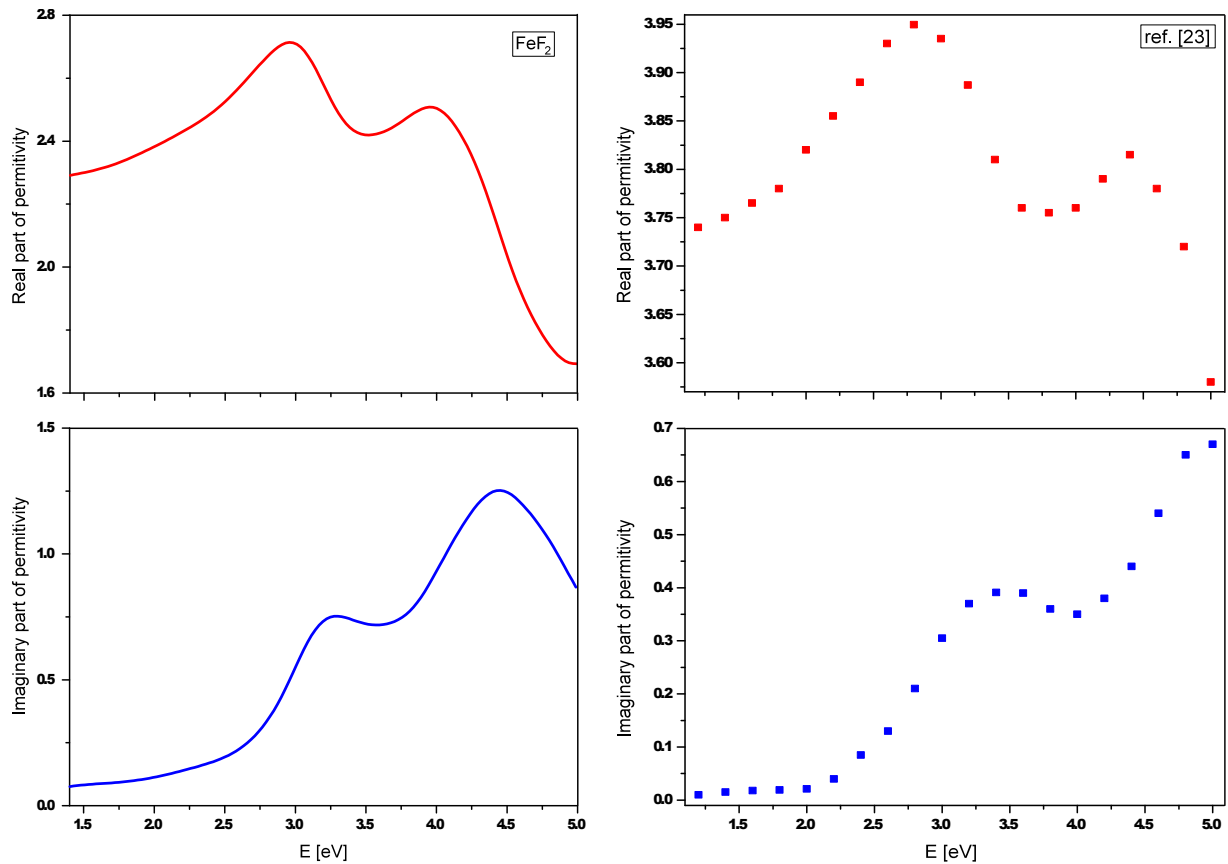}
\caption{On the left, optical spectra of FeF$_2$ (relative permittivity)
  calculated for $U=0.1$~Ry. Agreement with experimental data on
  the right (based on Fig.~4 of Pi\v stora~et~al.~[\onlinecite{Pistora:2010_a}])
  is striking, despite the fact that these measurements were taken
  above $T_N$.}
\label{fig-04}
\end{figure*}

The experimental work in question \cite{Pistora:2010_a} concerns 
room-temperature ellipsometry of FeF$_2$ layers. The gap inferred 
in Fig.~4 of that paper is certainly smaller than in
Ref.~[\onlinecite{Santos-Ortiz:2013_a}] and, moreover, it turns out that
the actual theoretical gap may be even smaller because of suppressed
transitions from the $d$-band directly below $E_F$ to the other
low-lying bands of group B and the ``conduction band'' (as defined in
Fig.~\ref{fig-01}). In fact, it is remarkable how similar are
theoretically calculated optical spectra for $U$ as small as 0.1~Ry
(shown in Fig.~\ref{fig-04}) to the experimental data \cite{Pistora:2010_a}
mentioned above. Given that optical gap at low temperatures will
probably be larger, we opted for showing band structure with
$U=0.2$~Ry in Fig.~\ref{fig-02}. Calculated magnetic moment, which is
smaller than the one experimentally determined (see
Tab.~\ref{tab-01}), also suggests that this choice of $U$ may be
better.


Both for FeF$_2$ and CoF$_2$, we chose rather small values of $U$ 
to have gaps around 2~eV in Fig.~\ref{fig-02}. This choice is
arbitrary and since measurements of structural parameters and/or
magnetic moment provide only a relatively benevolent
test \cite{Yang:2012_a} on the values of $U$ within usual \textit{ab initio}
calculations, not much progress can be expected here until
low-temperature optical measurements in a broad spectral range are
available. We show an example of such spectra (for CoF$_2$) in
Fig.~\ref{fig-06}: There is an abundance of spectral features that
could be tested against experiments. In fact, absorption edge around 0.8~eV
was measured \cite{Eremenko:1964_a} {\em at low temperature} in CoF$_2$, 
which would suggest very small value of $U$. This spectral feature is 
associated with a relatively narrow band whose origin could be in
interband transitions (see the inset of Fig.~\ref{fig-06}) but also in 
electron-phonon interactions \cite{Young:1969_a}, which would, in turn,
indicate a much larger gap. Nevertheless, large values of the Hubbard
parameter (like $U=7$~eV taken from cobalt
oxide \cite{Thunstrom:2012_a}) seem in contradiction with relatively
small MCA as explained in Sec.~III.

\begin{figure}
\includegraphics[scale=0.3]{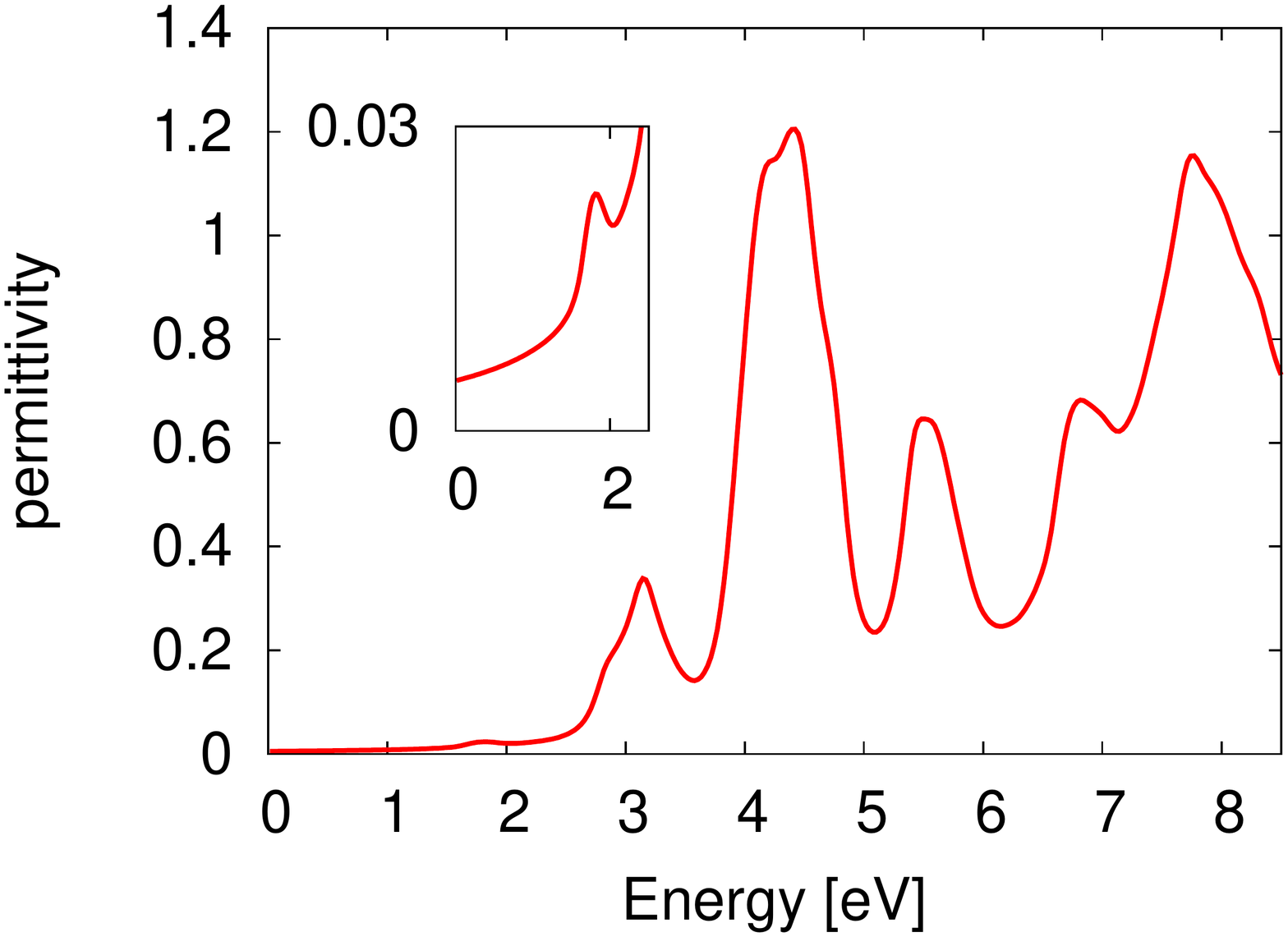}
\caption{Calculated imaginary part of permittivity, which is
  proportional to absorption, for CoF$_2$ ($U=0.1$~Ry). The inset
  shows a feature below 2~eV, which could be related to the observed
  narrow band \cite{Eremenko:1964_a}. }
\label{fig-06}
\end{figure}

Finally, NiF$_2$ again retains a gap even for vanishing $U$. Under the
action of the crystal field, the upper quintuplet of $d$-bands splits
into a doublet and a triplet. The former remains completely
depopulated and is separated by $\approx 1.5$~eV from the occupied
three bands forming bands of the group A (see the sketch in
Fig.~\ref{fig-01}). Similar to MnF$_2$, the effect of Hubbard $U$ is
to increase the gap size by moving the group A and B bands away from
each other. The small values of calculated TM magnetic moment in
Tab.~\ref{tab-01}, however, suggest that, similar to MnF$_2$, using
non-zero $U$ may be reasonable. To conclude this section, we stress
that in spite of uncertainty about what value of $U$ may lead to the
best description of the actual system, the qualitative character of the band
structure of all four compounds is clear: larger values of $U$ push
the bands in group B higher and make the optical gap larger. 
We believe that combining data from several different 
experimental sources provides a solid basis for band-structure validation 
and, to this end, we put emphasis on optical spectra in this work.

\section{Magnetocrystalline anisotropy}

Based on sufficiently accurately calculated electronic bands (with
effects of spin-orbit interaction included \cite{explSO}), total energy 
in the in-plane ($E_\parallel$) and out-of-plane magnetic
configurations ($E_\perp$) can be calculated. Sublattice magnetization
$M=S\mu_B/V$ (where $V$ is the volume of unit cell which contains one
TM atom of each sublattice) can then be used to obtain
$B_a=(E_\parallel-E_\perp)/M$ and also, using Eq.~(\ref{eq-02}),
$B_e=NJS^2/MV$. The central question now is how the total energies
depend on $U$: other quantities such as optimal structure parameters
(lattice constants) or TM magnetic moment do \cite{Yang:2012_a}, and it
is not {\em a priori} clear how sensitive the MCA is to the variation of $U$.

\begin{figure}
\includegraphics[scale=0.3]{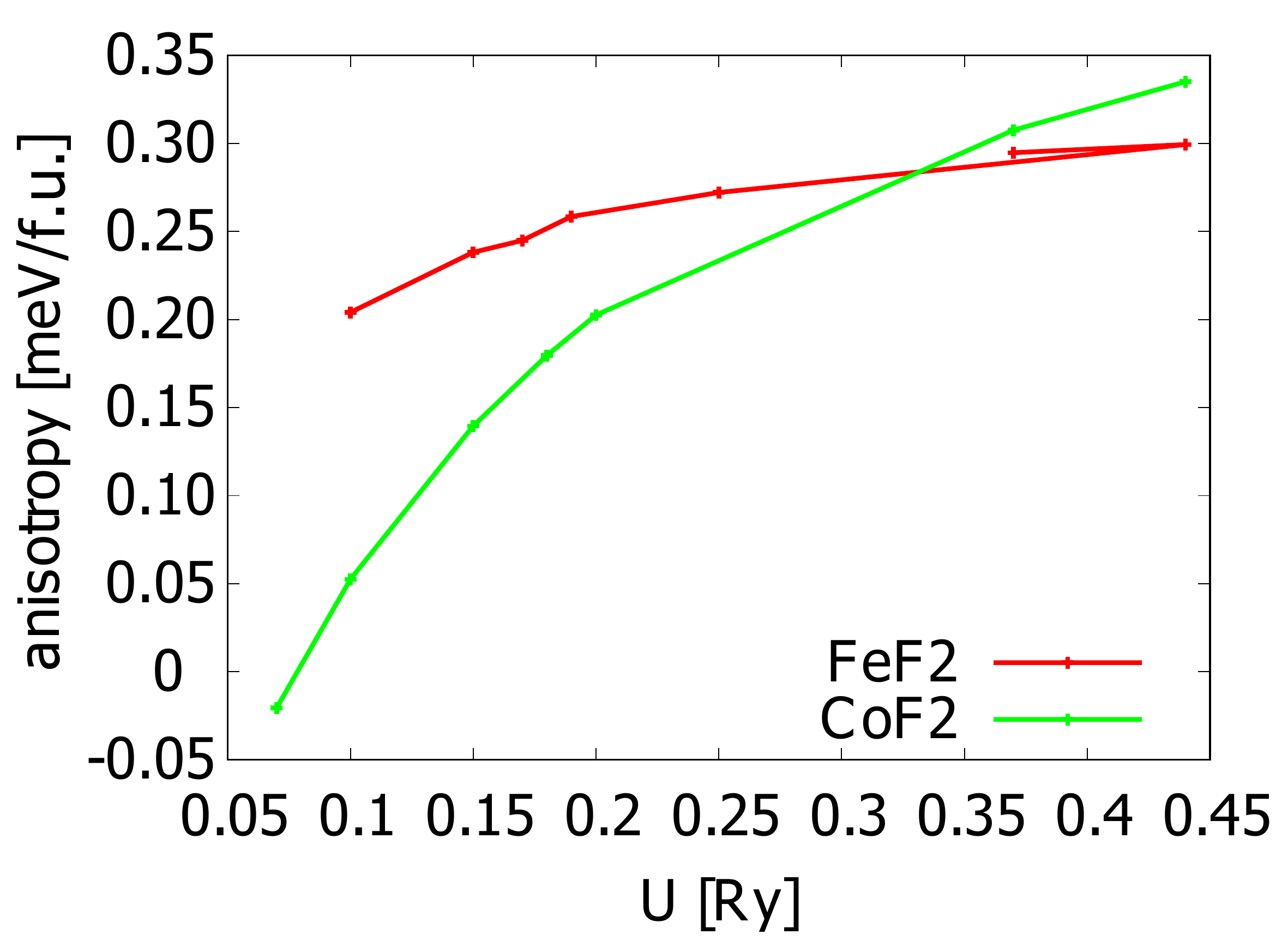}
\caption{Calculated MCA 
  for FeF$_2$ and CoF$_2$ as a function of the Hubbard parameter $U$.}
\label{fig-05}
\end{figure}

\begin{figure*}
\includegraphics[scale=0.25]{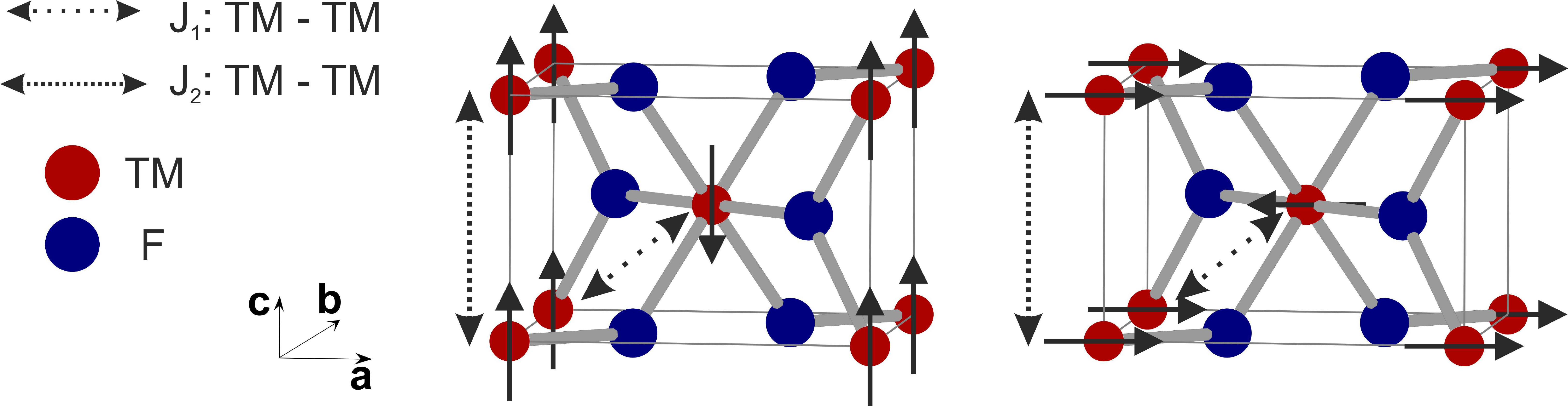}
\caption{Crystal structure (rutile) applies to all four difluorides
  under study. Magnetic structure on the left corresponds to
  orientation along the easy axis (except for NiF$_2$) and we denote its energy
  by $E_\parallel$. Magnetic structure on the right is defined to have
  energy $E_\perp$.}
\label{fig-08}
\end{figure*}

Keeping in mind that MCA {\em must} be very small for MnF$_2$ (recall the
argument of half-filled $d$-shell), there is no need to investigate its
dependence on $U$. The more-than-satisfactory agreement between
calculated and measured $B_{sf}$ in Tab.~\ref{tab-01} (the former one
being larger by 29\%) relies on the dipolar term, which is not as
difficult to evaluate. Regarding the quantitative agreement
between calculations and experiment, we should once again stress that
the limiting factor is now probably the estimate of $B_e$, based on
$T_N$. The situation is different with FeF$_2$: the value
$B_a^{(1)}=2.3~T$ in Tab.~\ref{tab-01} corresponds to calculations
with $U=0.2$~Ry.  Now, as Fig.~\ref{fig-05} shows, while the MCA does
depend on $U$, the variation is moderate. It is possible to conclude
that for FeF$_2$, magnetic anisotropies can be fairly well predicted
theoretically (Tab.~\ref{tab-01} shows that theoretical estimate of
$B_{sf}$ is only about 17\% lower than the measured value). Considering
the fact that sometimes \cite{Lopez-Moreno:2012_a} the \textit{ab initio} calculations
are extended to include another Hubbard-like parameter $J$, we also 
calculated MCA for $U=0.44$~Ry and $J=0.07$~Ry and found it to be somewhat
smaller than what would correspond to $U_{\mathrm{eff}}\equiv U-J=0.37$~Ry.
This further highlights the limits of quantitative predictions of MCA
based on \textit{ab initio} calculations.

As an alternative to \textit{ab initio} calculations, we note that
the sign and, to some extent, also the order of magnitude of MCA in
FeF$_2$ can be deduced from the single-ion model \cite{Hutchings:1970_a}.
Orbital multiplet of the Fe$^{2+}$ ion ($L=2$) is fully split by the
crystal field in the rutile structure and the action of spin-orbit
interaction $H_{SO}$ on the lowest (non-degenerate) level can be
written in terms of a spin $S=2$ Hamiltonian, $H_s=DS_z^2$. Corrections
to this form of $H_s$ are small \cite{Hutchings:1970_a}, derivation
of this result is explained below when we discuss CoF$_2$; note that
the argument in Eq.~(\ref{eq-06}) explains the negative
sign of $D$ as a consequence of level repulsion. Exchange
splitting $\beta$ oriented along the direction of the N\'eel vector
$\hat{e}_L$, combined with $H_s$, leads to a simple model exhibiting
MCA: $H_s+\beta \vec{S}\cdot\hat{e}_L$. The
lowest energy (with respect to spin) for $\hat{e}_L\parallel \hat{z}$
and $\hat{e}_L\parallel \hat{x}$, respectively, is thus obtained by
diagonalizing
\begin{equation}\label{eq-04}
  DS_z^2+\beta S_z\quad\mbox{and } DS_z^2+\beta S_x,
\end{equation}
which yields $4D-2\beta$ and $D-2\beta$ for the lowest eigenvalue
in the $\beta\gg D$ limit. Given $D<0$, the former direction
is preferred implying uniaxial anisotropy.
The values of $D$ (around 1~meV) determined by various experimental
techniques \cite{Lines:1967_a} are consistent, yet not quite in agreement with,
calculated and measured magnetic anisotropy of FeF$_2$. 

A very different situation is found with CoF$_2$. For small values of
$U$, MCA even changes sign (see again Fig.~\ref{fig-05}) and if we
take the experimental value of $B_{sf}$ in Tab.~\ref{tab-01} as a
means to estimate $B_a$ and, once the dipolar term has been
subtracted, also $B_a^{(1)}$, we find that the MCA changes with $U$
rapidly around the corresponding value (0.05~meV per formula unit). Hence
the conclusion, at minimum, that it is not possible to rely on
theoretical calculations of MCA in this case, unless some additional
guidance is provided. Moreover, values of $U$ that produce MCA
of this size are rather small (below 0.1~Ry) while more
commonly \cite{Thunstrom:2012_a} larger values are used. It should be
noted, however, that CoF$_2$ seems to be anomalous within the series
of four  materials considered in this paper (contrary to the other three, it 
has a significantly lower $T_N$) and it is possible that the estimate of
$B_e$ in Tab.~\ref{tab-01} is too large. This would allow for larger
$B_a^{(1)}$ and, in the spirit of  Fig.~\ref{fig-05}, for larger
values of $U$ as well. Reliable experimental determination of optical
gap (at low temperatures) could resolve this issue.

Analysis based on the single-ion model\cite{Ishikawa:1970_a} for Co$^{2+}$
leads quantitatively to an even worse estimate of MCA than for FeF$_2$ but
still predicts the correct sign and also the negligible in-plane
anisotropies. The orbital multiplet ($L=3$) is now split by octahedral
crystal field and the lowest lying $\Gamma_4$ triplet is further split
by \cite{Kamimura:1963_a} $\Delta\approx 0.1$~eV into a ground state
doublet and an excited state (singlet
$|L_z\rangle=|0\rangle$; in the following, we will use this notation
for the orbital part of wave functions). Rhombohedral crystal field
lifts the degeneracy of the doublet, producing states
\begin{equation}\label{eq-05}
  |a\rangle=\frac{\sqrt5}4|\!-\!3\rangle+\frac{\sqrt3}4|\!-\!1\rangle+
            \frac{\sqrt3}4|1\rangle+\frac{\sqrt5}4|3\rangle
\end{equation}
$$
  |b\rangle=\frac{\sqrt5}4|\!-\!3\rangle-\frac{\sqrt3}4|\!-\!1\rangle+
            \frac{\sqrt3}4|1\rangle-\frac{\sqrt5}4|3\rangle
$$
whose energy splitting $E_b-E_a$ is a fraction\cite{Kamimura:1963_a}
of $\Delta$. The perturbative action of
$H_{SO}=\lambda \vec{L}\cdot\vec{S}$ on the lower state can now be evaluated
to the second order in spin-orbit interaction $\lambda$. Provided we
neglect coupling to the $|L_z\rangle=|0\rangle$ state, we obtain
\begin{equation}\label{eq-06}
  \lambda^2
  \frac{\langle a|\vec{L}\cdot\vec{S}|b\rangle
        \langle b|\vec{L}\cdot\vec{S}|a\rangle}{E_a-E_b}=
  \frac{\frac94 \lambda^2}{E_a-E_b}S_z^2 \equiv DS_z^2
\end{equation}
because $\vec{L}\cdot\vec{S}=L_zS_z+\frac12(L_+S_-+L_-S_+)$ and,
given in Eq.~(\ref{eq-05}), the matrix elements of the raising (lowering)
operators $L_+$ ($L_-$) vanish. This construction predicts $D<0$ by
virtue of $E_a<E_b$ but, quantitatively, it implies a larger MCA than
for FeF$_2$ since
both $\lambda$ is larger (for CoF$_2$) and the energy splitting of the
lowest two states smaller. The absent in-plane anisotropy amounts to
$H_s$, containing no $S_x$, $S_y$ operators and this, in turn, is a
consequence of $\langle a|L_\pm|b\rangle=0$. Perturbative coupling to
the $|L_z\rangle=|0\rangle$ singlet will introduce the $S_{x,y}$ terms
to $H_s$, however, their coefficients will be small ($\Delta\gg |E_a-E_b|$).

Concerning the quantitative disagreement between the single-ion model
for CoF$_2$ and $E_\parallel-E_\perp$ calculated by \textit{ab initio}, a more
advanced approach seems necessary such as some kind of cluster model,
e.g., FeF$_6$, constructed along the lines of
Ref.~[\onlinecite{Subramanian:2013_a}] where a model of MnAs$_4$ cluster
was used to explain certain magnetic anisotropy terms in (Ga,Mn)As
dilute magnetic semiconductor. Such an attempt to make sense of the \textit{ab
initio} calculations is nevertheless clearly beyond the scope of this
paper. On the other hand, the single-ion model is successful in case
of FeF$_2$ and also \cite{sim-MnF2} MnF$_2$.

\section{Conclusion}

Magnetic anisotropies of MnF$_2$, FeF$_2$, CoF$_2$, and NiF$_2$ have
been investigated theoretically and it was found that, with the exception
of CoF$_2$, \textit{ab initio} calculations described in Appendix~C lead to
reliable results. For comparison to experiments, we used well-established spin-flop
 measurements (spin-flop field $B_{sf}$, see Tab.~\ref{tab-01}).
Regarding CoF$_2$, we conclude that while the calculations are
consistent with experimentally determined $B_{sf}$, the
MCA depends too sensitively on Hubbard
parameter $U$ so that quantitative prediction is impossible, without
knowing in advance what the correct result is.

We pointed out that band structures should be validated, for example,
through optical measurements, before using them for further
calculations. It would be desirable to perform such low-temperature
measurements for all four compounds and determine the optical
gap. This would afford greater confidence in the values of $U$ used in
\textit{ab initio} calculations.

\section*{Acknowledgements}

It is our pleasure to express gratitude to Pavel Nov\'ak for his
guidance in crystal field theories and literature dating back deep
into the 20th century.
We acknowledge support from National Grid Infrastructure MetaCentrum
provided under the programme ``Projects of Large Research, Development,
and Innovations Infrastructures'' (CESNET LM2015042) and also discussions
with Ond\v rej \v Sipr, Jind\v rich Koloren\v c, and Gerhard Fecher, and also 
funding via Contract No. 15-13436S (GA\v CR) and EU FET Open RIA Grant no. 766566 (ASPIN).
C.A.C. acknowledges the extended financial support by the ERDF, Project NanoCent 
No. CZ.02.1.01/0.0/0.0/15\_003/0000485.

\begin{appendix}

\section{Dipolar interactions}

Dipolar magnetic energy (per unit cell) of an (infinite) lattice of
magnetic moments is $E=-\frac12 \sum_j \vec{B}_j\cdot\vec{\mu}_j$ where the
sum goes over all magnetic moments $\vec{\mu}_j$ in the unit
cell. Magnetic field generated, at the position of given $\vec{\mu}_j$,
by all other magnetic moments is
\begin{equation}\label{eq-03}
  \vec{B}_j=\frac{\mu_0}{4\pi} \sum_i 
\frac{3(\hat{\mu}_i\cdot\hat{r}_{ij})\vec{r}_{ij}-\vec{\mu}_i}{|\vec{r}_{ij}|^3}
\end{equation}
where $\vec{r}_{ij}$ is the relative position of $\vec{\mu}_i$
with respect to $\vec{\mu}_j$. The dipolar magnetic energy depends on the
orientation of the magnetic moments; values labeled ``dipolar term'' in
Tab.~\ref{tab-01} are $E_\perp-E_\parallel$ recalculated into field
using the same procedure as for MCA (see Sec.~III). Magnetic moments
$|\vec{\mu}_j|$ used in Eq.~(\ref{eq-03}) were taken from
experiments \cite{Strempfer:2004_a}, as given in Tab.~\ref{tab-01}.

Dipolar interactions do not contribute to MA in cubic lattices while
they may even constitute its dominant source if the high symmetry is
broken (or completely absent).  To explain qualitatively the effect of
the broken symmetry, we consider a five-atom cluster (magnetic sublattice
A atom located at the center of coordinate system and four atoms of
magnetic sublattice B located at $(\pm a,0)$ and $(0,\pm b)$ with
strictly antiparallel magnetic moments) and calculate the energy of
the four B atoms in the dipolar field $\vec{B}_A$ implied by
Eq.~(\ref{eq-03}). This energy, $E(\phi)$, depends in general on the magnetic
moment orientation $(\sin\phi,\cos\phi)$. For $a/b=1$, however,
$E(\phi)$ is constant owing to $\sin^2\phi+\cos^2\phi$ being
independent on $\phi$. Once the symmetry is broken ($a\not=b$), the
configuration with moments parallel to $x$ ($\phi=\pi/2$) ceases to
have the same dipolar energy as the $\phi=0$ case, the ratio of the respective
energies being $(4-2a^3/b^3)/(4a^3/b^3-2)\not=1$.

\section{Structural parameters}

Crystal structure with two orientations of magnetic moments is shown
in Fig.~\ref{fig-08}. Antiferromagnetic ($J_1$) and ferromagnetic
($J_2$) interactions between nearby magnetic moments are highlighted,
the coupling $J$ discussed in Sec.~III is a weighted average of them.
Lattice constants and atom positions (taken from
Ref.~[\onlinecite{Baur:1971_a}]) are given in Tab.~\ref{tab-02}.
It is important to note that, upon introducing the magnetic order,
the space group of the crystal structure is modified from the tetragonal
$P4_2/mnm$ to the orthorhombic $Cmmm$ because the TM atoms sitting at
$(0,0,0)$ and $(1/2,1/2,1/2)$ are no longer equivalent by symmetry, as
their spins are antiparallel (Fig.~\ref{fig-08}). In our calculations,
we used muffin-tin radii ($R_{MT}$) as shown in Tab.~\ref{tab-03} and all data
shown in this paper are based on $Rk_{max}=7$.

\begin{table}
\caption{Crystal structure of MnF$_2$, FeF$_2$, CoF$_2$, and NiF$_2$
  \cite{Baur:1971_a}. Crystal system is tetragonal, space group
  $P4_{2}/mnm$ No. 136 for all four compounds.}
\label{t-c-struct}
\label{tab-02}
\begin{tabular}{lc} \hline
\multicolumn{2}{c}{\bf Chemical formula MnF$_2$}         \\
a, c (\AA)                    & 4.8738(1), 3.3107(1) \\
V (\AA$^3$)                   & 78.642(3) \\
\multicolumn{2}{c}{atomic positions} \\
Mn1                           & (0,0,0) \\
Mn2                           & (0.5,0.5,0.5) \\
F1                            & (0.3053(12),0.3053(12),0) \\ 
F2                            & (0.8053(12),0.1947(12),0.5) \\[3mm]
\multicolumn{2}{c}{\bf Chemical formula FeF$_2$}         \\
a, c (\AA)                    & 4.6945(4), 3.3097(1) \\
V (\AA$^3$)                   & 72.940(9) \\
\multicolumn{2}{c}{atomic positions} \\
Fe1                           & (0,0,0) \\
Fe2                           & (0.5,0.5,0.5) \\
F1                            & (0.3010(8),0.3010(8),0) \\ 
F2                            & (0.8010(8),0.1990(8),0.5) \\[3mm]
\multicolumn{2}{c}{\bf Chemical formula CoF$_2$}         \\
a, c (\AA)                    & 4.6954(4), 3.1774(4) \\
V (\AA$^3$)                   & 70.051(12) \\
\multicolumn{2}{c}{atomic positions} \\
Co1                           & (0,0,0) \\
Co2                           & (0.5,0.5,0.5) \\
F1                            & (0.3052(8),0.3052(8),0) \\ 
F2                            & (0.8052(8),0.1948(8),0.5) \\[3mm]
\multicolumn{2}{c}{\bf Chemical formula NiF$_2$}         \\
Crystal system, space group   & Tetragonal, $P4_{2}/mnm$ No. 136 \\
a, c (\AA)                    & 4.6498(3), 3.0838(1) \\
V (\AA$^3$)                   & 66.674(6) \\
\multicolumn{2}{c}{atomic positions} \\
Ni1                           & (0,0,0) \\
Ni2                           & (0.5,0.5,0.5) \\
F1                            & (0.3012(13),0.3012(13),0) \\ 
F2                            & (0.8012(13),0.1988(13),0.5) \\
\end{tabular}
\end{table}

\begin{table}
  \caption{$R_{MT}$ in Bohr radii for individual atoms used in our
    calculations.}
  \begin{tabular}{ccc|ccc}
    Mn, F & 2.08, 1.88 &&& Co, F & 1.97, 1.78 \\
    Fe, F & 1.97, 1.78 &&& Ni, F & 1.95, 1.77 
  \end{tabular}
  \label{tab-03}
\end{table}

\section{Electronic structure calculations}

For our density functional theory (DFT) calculations, we use the
linearized augmented plane wave method \cite{Blaha:2001}, with added
orbital-dependent correction (so called DFT+U). The chosen orbitals
for this Hubbard-like term are the $3d$-states of the TM,
 scheme of Ref.~[\onlinecite{Anisimov:1993_a}] is used to avoid
double counting and GGA to the density functional is employed. Scalar
relativistic approach to the spin-orbit interaction is taken from the
very beginning of our calculations. 
Convergence with respect to the number of points in the
$k$-space ($n_k$) is achieved already around $n_k=10000$, meaning that
energies $E_\parallel$ and $E_\perp$ are converged to several $\mu$Ry
while their difference is about an order of magnitude larger. Within
this precision, we find no significant difference between in-plane
directions, which agrees, assuming the single-ion model to be valid,
with the perturbative argument [see Eq.~(\ref{eq-05}) and below].

\begin{figure}
\includegraphics[scale=0.45]{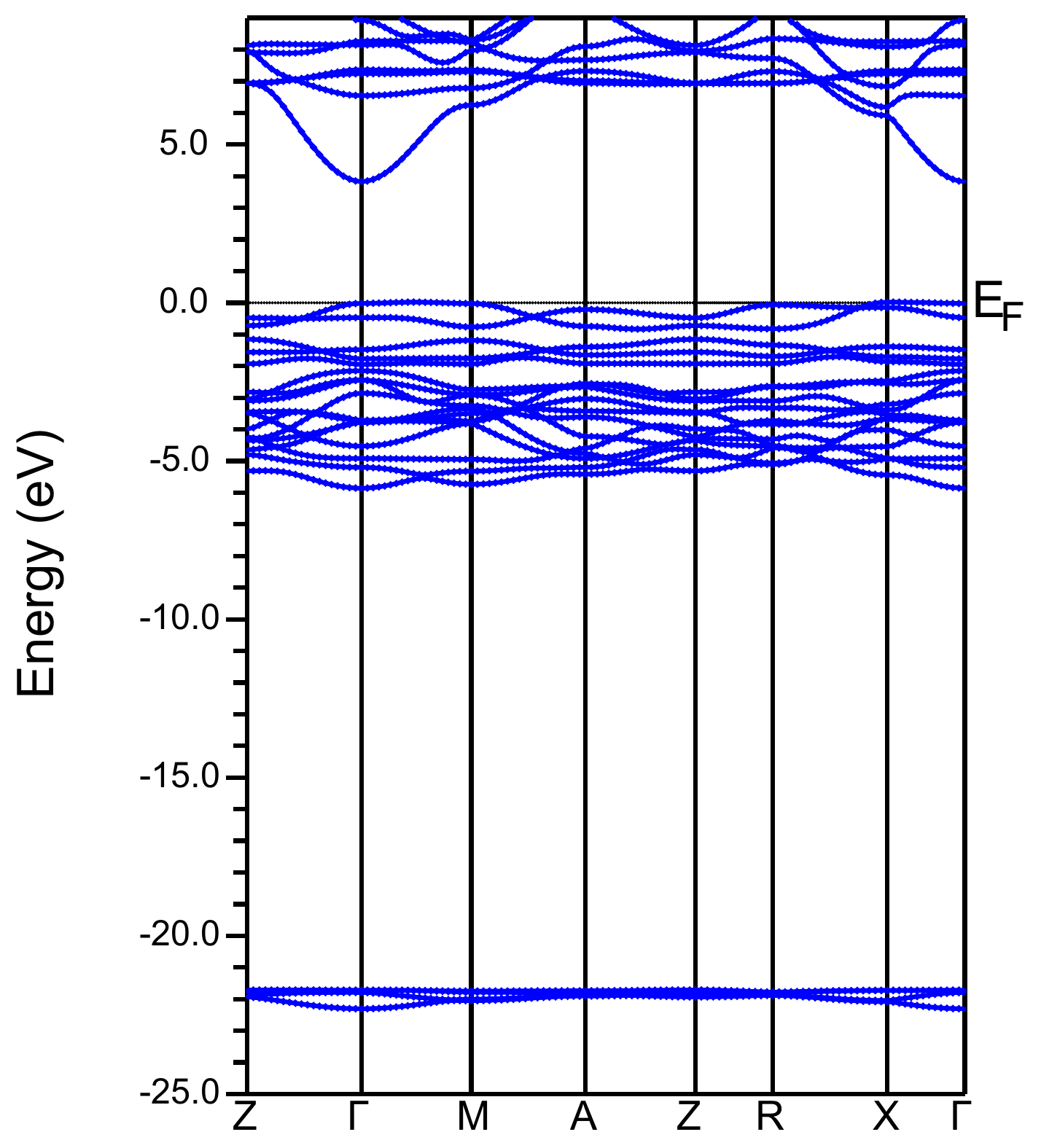}
\caption{Band structure of MnF$_2$ calculated with large $U$
  corresponding to Ref.~[\onlinecite{Stavrou:2016_a}] (0.43~Ry).}
\label{fig-07}
\end{figure}

As described in the main text and Fig.~\ref{fig-01}, the main effect
of increasing $U$ is to push the group B $d$-bands away from the lower
quintuplet of the $d$-states and, if present (as for FeF$_2$, CoF$_2$,
and NiF$_2$), also from the group A $d$-bands. This can be seen by comparing
Fig.~\ref{fig-07} (exemplifying the effect of large $U$) to the
leftmost panel of Fig.~\ref{fig-02}. As a side remark, we note that
the position of the low-lying fluorine $2s$ states also depends on the
TM ion type.

Figure~\ref{fig-07} also clearly shows the ``conduction band'' (lowest
lying unoccupied parabolic band). Its effective mass is moderately
anisotropic and smaller than the free electron rest mass $m_0$; wave functions
of this band are largely localized in the interstitial space.
Averaged over directions, we find $m_{\mathrm{eff}}/m_0$ about 0.22
for MnF$_2$, 0.25 for FeF$_2$, 0.51 for CoF$_2$, and 0.36 for NiF$_2$.

\end{appendix}


\begin{thebibliography}{99}


\bibitem{Baltz:2017_a} V. Baltz et al., Rev. Mod. Phys. 90, 015005-1 (2018);
  T. Jungwirth et al., Nat. Phys. 14, 200 (2018).
        
\bibitem{Johansen:2017_a} {\O}. Johansen and A. Brataas,
   Phys. Rev. B 95, 220408 (2017). 

\bibitem{Rezende:2016_a} S. M. Rezende et al.,
   Phys. Rev. B 93, 014425 (2016).

\bibitem{Wu:2016_a} S.M. Wu et al., 
Phys. Rev. Lett. 116, 097204 (2016).

\bibitem{Moriyama:2015_a} T. Moriyama et al.,
 Appl. Phys. Lett. 106, 162406 (2015).

\bibitem{Gulbrandsen:2018_a} S.A. Gulbrandsen and A. Brataas,
   Phys. Rev. B 97, 054409 (2018).
  
\bibitem{Bogdanov:2007_a} A.N. Bogdanov et al., Phys. Rev. B 75, 094425 (2007). 
        
\bibitem{Strempfer:2004_a} J. Strempfer et al., Phys. Rev. B 69, 014417 (2004).
        
\bibitem{Novak:2006_a} P. Nov\'ak et al.,
   Phys. Stat. Sol. (B) 243, 563 (2006).
        
\bibitem{Carrico:1994_a} A.S. Carri\c{c}o et al.,
   Phys. Rev. B 50, 13453 (1994).
        
\bibitem{Jacobs:1961_a} I.S. Jacobs,
   J. Appl. Phys. 32, s61 (1961).
        
\bibitem{Jaccarino:1983_a} V. Jaccarino et al.,
 J. Magn. Magn. Mater. 31-34, 1117 (1983).
        
\bibitem{Nagamiya:1955_a} T. Nagamiya et al., 
   Adv. Phys. 4, 1 (1955).

\bibitem{Balkanski:1964_a} M. Balkanski et al.,
   J. Chem. Phys. 40, 1897 (1964).
	  
\bibitem{Shi:2004_a} H. Shi et al.,
 Phys. Rev. B 69, 214416 (2004).

\bibitem{Dufek:1993_a} P. Dufek et al.,
   Phys. Rev. B 48, 12672 (1993).
                
\bibitem{Dufek:1994_a} P. Dufek et al.,
   Phys. Rev. B 49, 10170 (1994).

\bibitem{Lopez-Moreno:2012_a} S. L\'opez--Moreno et al.,
 Phys. Rev. B 85, 134110 (2012).

\bibitem{Hutchings:1970_a} M.T. Hutchings et al.,
  J Phys C: Sol St Phys 3, 307 (1970).
	
	\bibitem{Lines:1967_a} M.E.~Lines,
  Phys. Rev. 156, 543 (1967).
  
\bibitem{Ishikawa:1970_a} A. Ishikawa and T. Moriya,
  J. Phys. Soc. Jpn. 30, 117 (1970).

\bibitem{Barreda-Argueso:2013_a} J.A. Barreda--Arg\"ueso et al.,
   Phys. Rev. B 88, 214108 (2013).
        
\bibitem{Valerio:1995_a} G. Valerio et al.,
   Phys. Rev. B 52, 2422 (1995).
 
\bibitem{Santos-Ortiz:2013_a} R. Santos-Ortiz et al.,
  Appl. Mater. Interfaces 5, 2387 (2013).
		
\bibitem{Thunstrom:2012_a} P. Thunstr\"om et al.,
  Phys. Rev. Lett. 109, 186401 (2012).  

\bibitem{Pistora:2010_a} J. Pi\v{s}tora et al.,
  J. Phys. D:Appl. Phys. 43, 155301 (2010).
	
\bibitem{Yang:2012_a} Z. Yang et al.,
   Trans. Nonferrous Met. Soc. China 22, 386 (2012). 
	
\bibitem{Eremenko:1964_a} V.V. Eremenko and A.I. Zvyagin,
  Sov. Phys.‐Solid State 6, 1013 (1964), in Russian: Fiz. Tverd. Tela 6, 1013 (1964).

\bibitem{Young:1969_a} P.A. Young,
  Thin Sol. Films 4, 25 (1969).
	
\bibitem{Kamimura:1963_a} H. Kamimura and Y. Tanabe,
  J Appl Phys 34, 1239 (1963).
	
\bibitem{Subramanian:2013_a} H.~Subramanian and J.E.~Han,
  J. Phys.: Cond. Matter 25, 206005 (2013).
	
\bibitem{sim-MnF2} Within the single-ion model, the ground state of
  a Mn$^{2+}$ ion is an orbital singlet according to the Hund's
  rule. Hence, the ground state cannot be split by crystal field and
  the only way for $H_{SO}$, defined above Eq.~(\ref{eq-05}), to
  come into play is to
  mix the $^6S$ singlet with an excited $^4P$ state. Energy of the
  latter is orders of magnitude larger than the crystal field
  splittings in FeF$_2$, CoF$_2$, and NiF$_2$ so that the perturbative
  effect of  $H_{SO}$ on the Mn$^{2+}$ ground state and hence MCA is
  small. 
    
\bibitem{Baur:1971_a} W.H. Baur and A.A. Khan,
   Acta Cryst. B 27, 2133 (1971).
	
\bibitem{Stavrou:2016_a} E. Stavrou et al.,
   Phys. Rev. B 93, 054101 (2016).
			
\bibitem{explEq2} This is a standard result of Weiss theory of
  magnetism [see, for instance, Eq.~(11) in Phys. Rev. 123, 425], where
  $H_i=-J\langle \vec{S}\rangle \cdot \vec{S}_i$ is assumed for
  interaction between a given localized magnetic moment $\vec{S}_i$ and a
  mean field $\langle \vec{S}\rangle$ produced by all other magnetic
  moments.

\bibitem{explMMMnF2} Note that the calculated value of magnetic moment
  in Tab.~\ref{tab-01} is significantly lower than the experimental
  value. This suggests that appreciable $U$ should in fact be used.

\bibitem{explSO} Band structure does not change when direction of
  magnetic moments is varied unless spin-orbit interaction is included
  in the DFT calculations.

\bibitem{Blaha:2001} P. Blaha et al., 
   {\em WIEN2k, An Augmented Plane Wave + Local Orbitals Program for
   Calculating Crystal Properties} (Karlheinz Schwarz, Techn. Universit\"at
   Wien, Austria, 2001).

\bibitem{Anisimov:1993_a} V.I. Anisimov et al.,
   Phys. Rev. B 48, 16929 (1993). 
  
\end{thebibliography}

\end{document}